\documentclass[X=12pt,oneside,pdflatex,sn-mathphys-num]{sn-jnl}
\makeatletter
\@twosidefalse  
\@mparswitchfalse  
\makeatother

\usepackage{graphicx}%
\usepackage{multirow}%
\usepackage{amsmath,amssymb,amsfonts}%
\usepackage{amsthm}%
\usepackage{mathrsfs}%
\usepackage[title]{appendix}%
\usepackage{xcolor}%
\usepackage{adjustbox}
\usepackage{textcomp}%
\usepackage{manyfoot}%
\usepackage{booktabs}%
\usepackage{algorithm}%
\usepackage{algorithmicx}%
\usepackage{algpseudocode}%
\usepackage{listings}%
\usepackage{hyperref}
\usepackage{makecell}
\usepackage{geometry}
\usepackage[utf8]{inputenc}
\usepackage{siunitx}
\usepackage{array}
\usepackage{caption}

\newcommand{\red}[1]{{\color{black} #1}}

\begin{document}


\title[Article Title]{Inferring Tight-Binding Hamiltonian Parameters from ARPES Spectra via Implicit Neural Representations}


\author[1]{\fnm{Yu} \sur{Zhang}}
\equalcont{These authors contributed equally to this work.}

\author[2,3]{\fnm{Yong} \sur{Zhong}}
\equalcont{These authors contributed equally to this work.}

\author[1]{\fnm{Nhat Huy} \sur{Tran}}

\author[1]{\fnm{Shuyi} \sur{Li}}

\author[3,4]{\fnm{Kyuho} \sur{Lee}}

\author[2,3]{\fnm{Yonghun} \sur{Lee}}

\author[2,3]{\fnm{Tiffany C.} \sur{Wang}}

\author[2,3]{\fnm{Harold Y.} \sur{Hwang}}

\author*[2,3,4]{\fnm{Zhi-Xun} \sur{Shen}}\email{zxshen@stanford.edu}

\author*[1]{\fnm{Chunjing} \sur{Jia}}\email{chunjing@phys.ufl.edu}

\affil[1]{\orgdiv{Department of Physics}, \orgname{University of Florida}, \orgaddress{\city{Gainesville}, \state{FL} \postcode{32611}, \country{United States}}}

\affil[2]{\orgdiv{Department of Applied Physics}, \orgname{Stanford University}, \orgaddress{\city{Stanford}, \state{CA} \postcode{94305}, \country{United States}}}

\affil[3]{\orgdiv{Stanford Institute for Materials and Energy Sciences}, \orgname{SLAC National Accelerator Laboratory}, \orgaddress{\city{Menlo Park},  \state{CA} \postcode{94025}, \country{United States}}}

\affil[4]{\orgdiv{Department of Physics}, \orgname{Stanford University}, \orgaddress{\city{Stanford}, \state{CA} \postcode{94305}, \country{United States}}}

\abstract{The emergent behavior of quantum materials is governed by their electronic structure, which can be experimentally investigated through angle-resolved photoemission spectroscopy (ARPES) and theoretically described by effective Hamiltonians. However, the quantitative extraction of Hamiltonian parameters from these high-dimensional spectra remains a significant challenge, currently relying on labor-intensive, expert-dependent analysis rather than standardized workflows. Here, we introduce a deep learning framework based on implicit neural representations to infer tight-binding Hamiltonian parameters directly from ARPES spectra. We demonstrate our approach on two classes of transition-metal oxides: perovskite nickelates and manganites, achieving good agreement with experimental Fermi surface topologies and energy-momentum dispersions. This work highlights the potential of deep learning tools to bridge the gap between theory and experiment, paving the way for high-throughput, autonomous discovery pipelines in quantum materials.}

\maketitle
\newpage
\section*{Introduction}
\label{Sec:Intro}

Artificial Intelligence (AI) has catalyzed a paradigm shift in the natural sciences, most dramatically in biology, where deep learning architectures such as AlphaFold have solved long-standing challenges in protein structure prediction and gene sequence analysis \cite{AlphaFold, Gene}. This breakthrough provides a compelling blueprint for the next frontier in materials science: the AI-driven discovery of quantum materials. By leveraging deep learning tools to decode materials' ``genome" — the underlying electronic structures, novel quantum states can be rationally designed to meet the demands of high-temperature superconductivity, exotic magnetism, and non-trivial topology.

Modern AI techniques have emerged as a transformative force in the theoretical prediction of electronic structures. At the density functional theory (DFT) level, recent developments include efficient mapping of potential energy surfaces via DeepMD~\cite{DeepMD,DeepMDv2} and general machine learning interatomic potentials (MLIPs) such as EquiformerV2 and MatterSim~\cite{EquiformerV2, mattersim}, as well as direct prediction of electronic band dispersions using frameworks like DeepH~\cite{DeepH,DeepHE3}. These pioneering approaches learn from  large datasets of first-principles DFT or $ab$ initio molecular dynamics calculations, enabling the simulation for greatly more complex systems with remarkable efficiency. Beyond interatomic potentials, neural-network quantum states (NQS) provide a powerful route to solving strongly correlated systems based on model Hamiltonians. By employing restricted Boltzmann machine (RBM)~\cite{Carleo2017} and Transformers-based~\cite{TransformerQSZhang, Sobral2025} architectures, open-source tools such as NetKet~\cite{NetKet} and DeepSolid~\cite{DeepSolid} capture strong electronic correlations and entanglement effects beyond the mean-field level. Recent developments have extended these capabilities to spectral functions~\cite{MendesSantos2023, Yoshioka2021, Douglas2021}, enabling investigation of both static and dynamical electronic properties.

Although rapid advances have been achieved in the theoretical frontier of electronic structure prediction, the ability to benchmark these results with experimental observations remains a critical bottleneck. 
Angle-resolved photoemission spectroscopy (ARPES) is the most direct experimental technique for probing the electronic structure of quantum materials \cite{ARPES}. By resolving both energy ($E$) and momentum ($\mathbf{k}$), it generates a four-dimensional dataset ($E, k_x, k_y, k_z$) that encapsulates the full electronic information. Despite these unparalleled capabilities, high-quality ARPES datasets are severely limited by practical barriers: experiments require access to large-scale synchrotron facilities and involve low-throughput acquisition cycles spanning tens of hours. Moreover, downstream analysis is labor-intensive and contingent upon expert bias. Therefore, the development of dedicated AI frameworks is essential to accelerate the analysis of high-dimensional spectral data. 

Deep learning has demonstrated increasingly important utilities across diverse spectroscopic techniques. It has been widely used to uncover crucial material properties through feature extraction \cite{Chen2021X-ray, Chitturi2023Neutron, Ziatdinov2022STM}, high-fidelity denoising \cite{Oppliger2024Denoising} and spectral prediction \cite{PhysRevLett.124.156401}. Within the specific domain of ARPES, nascent efforts have employed deep learning architectures to address specialized problems, such as probabilistic band retrieval \cite{Xian2023Band}, superconducting gap determination \cite{Chen2025SC}, and multidimensional data cleaning \cite{Kim2021Denoise}. Despite these successes, an intensity-based framework that infers the parameters of an effective Hamiltonian directly from the measured spectra, rather than from extracted peak positions, remains largely unexplored.
This objective is hindered by several challenges: the scarcity of standardized experimental repositories, the intrinsic complexity of many-body electronic interactions, and the experimental noise inherent to the measurement process.

Implicit neural representations (INRs) encode signals into the weights of neural networks that map coordinates, such as spatial positions or pixel locations, to values, such as surface distances or colors\cite{DeepSDF, mescheder2019occupancynetworkslearning3d}. INRs have been recently applied to neutron spectroscopy analysis\cite{SIREN_sqw}, where the continuity and differentiability of the representation make the inverse problem realizable: the physical parameters given as inputs can be inferred from the output signals by gradient-based optimization. As ARPES probes the electronic structure in energy-momentum space, the spectral intensity is naturally a function of the coordinates $(E, \mathbf{k})$, which makes INRs well suited to representing ARPES spectra.



In this work, we apply this idea to ARPES analysis and present an intensity-based framework that extracts tight-binding Hamiltonian parameters directly from experimental ARPES data. Specifically, we train an INR on a tight-binding model, so that it serves as a differentiable surrogate of the model. We use two real-world examples, perovskite nickelate and manganite, to demonstrate the power of deep learning methods for ARPES data interpretation. The algorithm delivers reconstructions that agree well with experimental spectra for both materials, demonstrating that, within an assumed tight-binding model, an ML‑based approach can automatically extract electronic parameters from high-dimensional ARPES data with high accuracy, computational efficiency and transferability. 

\section*{Results}
\label{Sec:Results}

The perovskite nickelate Nd$_{1-x}$Sr$_x$NiO$_3$ is a promising candidate for next-generation electronic devices, characterized by a sharp, tunable metal-insulator transition \cite{Medarde1997, Torrance1992, Catalano2018, Middey2016, Chakhalian2014}. Its low-energy electronic structure is primarily governed by the two $e_g$ orbitals of the $\mathrm{Ni}^{3+}$ ion, as shown in Figure 1{\bf a}. Recently, systematic ARPES measurements have succeeded in mapping the high-quality, four-dimensional electronic structure of Nd$_{1-x}$Sr$_x$NiO$_3$ with $x=0.175$ \cite{Yong_NdNiO3}. This rich experimental dataset makes perovskite nickelate an ideal model system for developing ML tools to automatically extract the essential electronic parameters underpinning its unique properties.

\begin{figure}[htbp]
    \centering
    \includegraphics[width=1.0\linewidth]{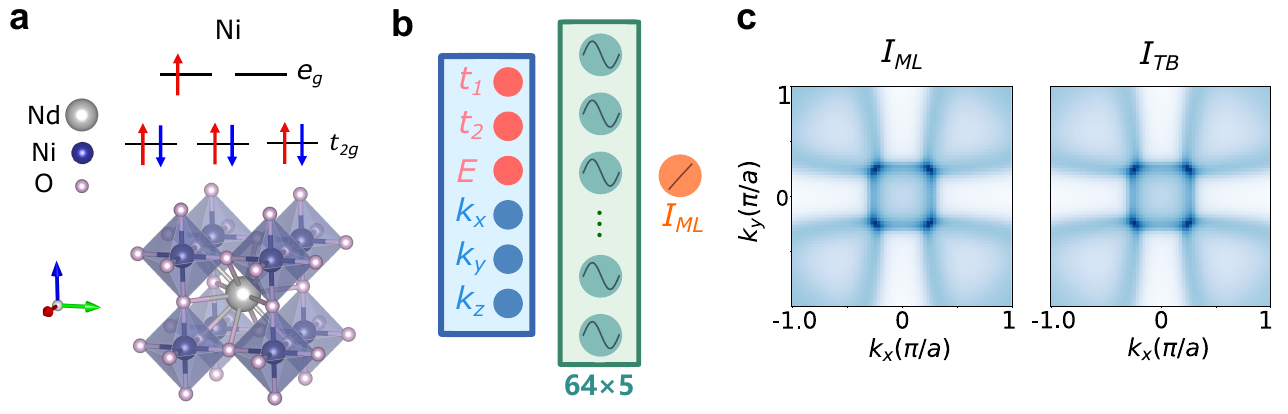}
    \caption{\textbf{a}. Crystal structure of Nd$_{1-x}$Sr$_x$NiO$_3$ with the corresponding electronic configuration of Ni$^{3+}$.  
    \textbf{b}. Architecture of the SIREN model: five hidden layers are placed between the inputs ($t_1$, $t_2$, the energy $E$, $k_x$, $k_y$ and $k_z$) and the output (spectral intensity $I_{ML}$).  
    \textbf{c}. Fermi surface obtained from the trained machine learning model (left) and from the tight-binding model (right).
    }
    \label{ML_model}
\end{figure}

\subsection*{Two-Band Tight-Binding Model}

To capture the essential low-energy physics of perovskite nickelate, we employ a tight-binding model to simulate its electronic structure. This model is designed to reproduce two critical features in the ARPES study: band dispersion and Fermi surface topology. The band dispersion, $E(\mathbf{k})$, describes the relationship between the energy of an electron and its momentum. The Fermi surface---the boundary in momentum space separating occupied from unoccupied states---dictates many of the key properties of the material. Based on the electronic configuration of the Ni$^{3+}$ ion, a two-band tight-binding Hamiltonian can provide an effective description of the band dispersion and the Fermi surface topology for Nd$_{1-x}$Sr$_x$NiO$_3$ system \cite{TBmodel}:
\begin{equation}
    H_{tb}=-\sum_{ij}t_{ij}^{ab}c_{ia}^{\dagger}c_{jb}-\mu
\end{equation}
Here, $i$, $j$ are site indices, $a, b = 1, 2$ are orbital indices corresponding to $d_{3z^2 - r^2}$ and $d_{x^2 - y^2}$ orbitals, and $\mu$ is the chemical potential which can be tuned by changing the doping levels. Only the nearest-neighbor hopping $t_1$ and next-nearest-neighbor hopping $t_2$ with $\sigma$-type bonding are considered in our simulations. 

 We can calculate ARPES spectral intensity $I_{TB}(E,\bold{k})$ through equations (2) and (3), where $\epsilon(\mathbf{k})$ represents the bare band, $\Sigma'(E)$ and $\Sigma''(E)$ are the real and imaginary parts of the self-energy, respectively, and $f(E)$ is the Fermi-Dirac function. The energy $E$ is measured relative to the Fermi level, with $E=0$ corresponding to the Fermi energy. Since we consider only occupied states at low temperatures, the Fermi–Dirac distribution is not explicitly included.
 The $k_z$ broadening is taken into account by applying a Lorentzian convolution with $\Delta k_z=0.2\frac{\pi}{a}$. The resulting spectral intensity varies with different choices of $(t_1, t_2, \mu)$.
\begin{equation}
    A(E,\bold{k})=\sum -\frac{1}{\pi}
    \frac{\Sigma^{''}(E)}{(E-\epsilon(\bold{k})-\Sigma^{'}(E))^2+(\Sigma^{''}(E))^{2}}
\end{equation}

\begin{equation}
    I_{TB}(E,k_x, k_y)=\int_{k_z^0-\frac{\Delta k_z}{2}}^{k_z^0+\frac{\Delta k_z}{2}}f(E)\frac{A(E,\bold{k})}{(k_z-k_z^0)^2+(\Delta k_z/2)^2}
\end{equation}

\subsection*{Implicit Neural Representations}
Conventional analytical approaches for extracting electronic parameters from ARPES data are often hindered by complex dependence on model parameters, the vast momentum-energy configurational space, and the presence of experimental noise. To address these challenges, we developed a ML framework that utilizes implicit neural representations to navigate the tight-binding parameter space. Our method, inspired by the work of Chitturi et al.  \cite{SIREN_sqw} for neutron spectroscopy analysis, provides a robust and high-fidelity alternative for parameter extraction. Specially, our model aims to obtain the tight-binding parameters ($t_1,t_2,\mu$) that best fit the experimental ARPES spectra.

To simulate the tight-binding model for different parameters, we implement a fully connected neural network with sinusoidal activation functions, specifically the sinusoidal implicit neural representation (SIREN) model, as shown in Figure 1{\bf b}. SIREN model can capture fine details for natural signals \cite{SIREN}. Considering the intrinsic three-dimensional electronic structure of Nd$_{1-x}$Sr$_x$NiO$_3$, we chose $(k_x, k_y, k_z, t_1, t_2, E)$ as inputs, and the ML model is trained to approximate the ARPES intensity $I_{ML}(E,\bold{k})$. While discrete data points are used for training, the ML model is capable of interpolation to provide a continuous representation.

Based on the well-trained and differentiable neural network, we optimize the unknown tight-binding parameters using a gradient-based optimization algorithm. More specifically, we use the Pearson correlation coefficient $r$\cite{SIREN_sqw}:
\begin{equation}
    r=\frac{Cov(y,y_{pred})}{\sigma_y \sigma_{y_{pred}}}
\end{equation}
as a metric to assess the similarity between $I_{ML}$ and $I_{TB}$ (Figure 1{\bf c}). We assume that these two values are linearly correlated. Since $r$ is invariant under linear transformation, we can avoid the normalization problem in this case. We define the loss function as $L = 1-r$, using the SIREN model as the surrogate for the tight-binding model. We minimize $L$ via the Adam optimizer\cite{Adam}, following the workflow as shown in Figure 2{\bf a}.


\begin{figure}[htbp]
  \centering
  \includegraphics[width=1.0\linewidth]{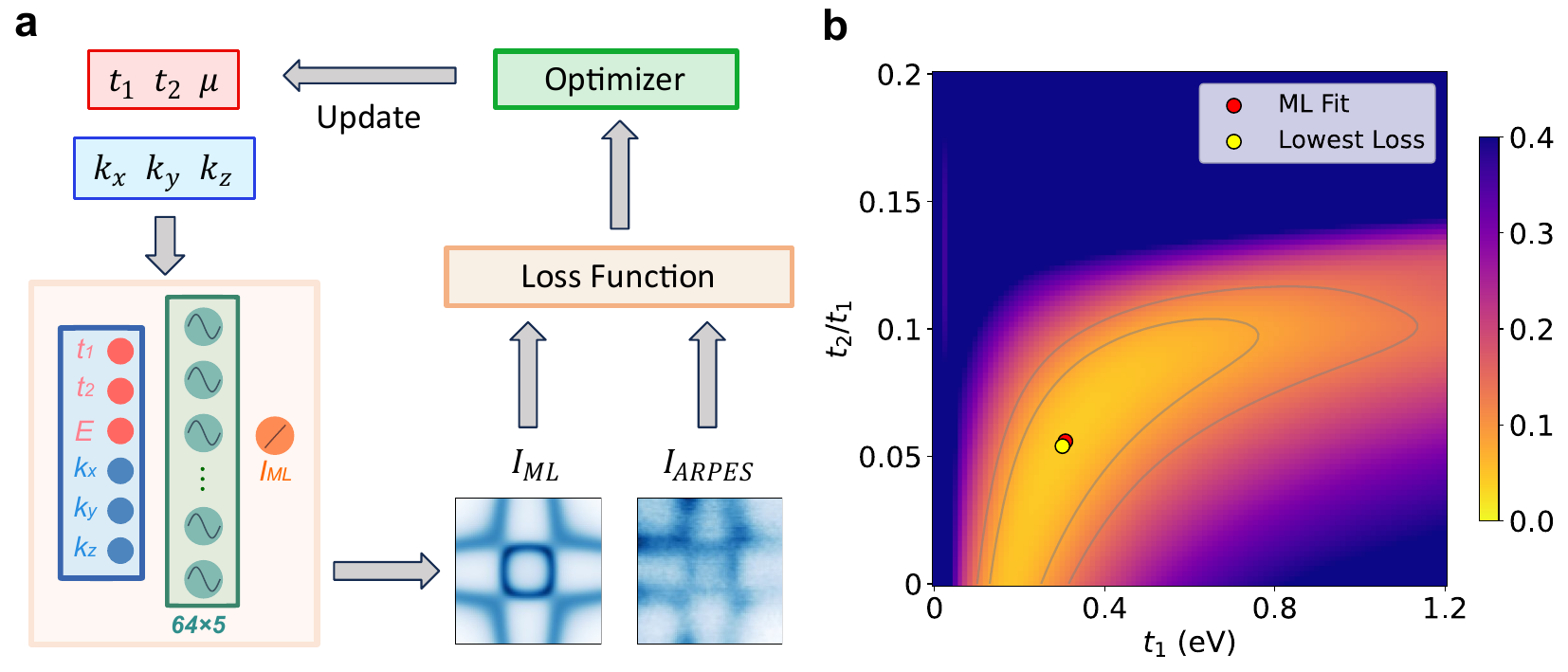}
  \caption{
    \textbf{a}. Our machine learning workflow to obtain the material parameters. An initial guess of the tight-binding parameters is fed into the well-trained SIREN model. The parameters are iteratively updated using the gradient $\frac{\partial L}{\partial t_i}$ until convergence.  
    \textbf{b}. Visualization of the loss landscape at $\mu = 1.95t_1$. A grid of 101 sampled values was employed for both $t_1$ and $t_2/t_1$. The color map represents the loss calculated by the trained ML model for the corresponding tight-binding parameters. Red and yellow points indicate the results from the ML fit method and grid search (101$\times$101) under specific $\mu$, respectively.
    }
  \label{flowchart}
\end{figure}

Figure 2{\bf b} shows that the tight-binding parameters, nearest neighbor hopping $t_1$ and the ratio of next-nearest-neighbor to nearest-neighbor hopping $t_2/t_1$, obtained from the gradient optimization method lie close to the global minimum of the loss landscape at fixed chemical potential $\mu = 1.95t_1$, providing a reliable approximation. The small discrepancies between the ML-obtained parameters and the point corresponding to the minimum loss can be attributed to two factors: (1) the evaluation is performed at a fixed $\mu$ level. (2) the discrete grid search, which may not capture the global minimum due to limited resolution.


\begin{figure}[H]
\centering
    \includegraphics[width=1.0\linewidth]{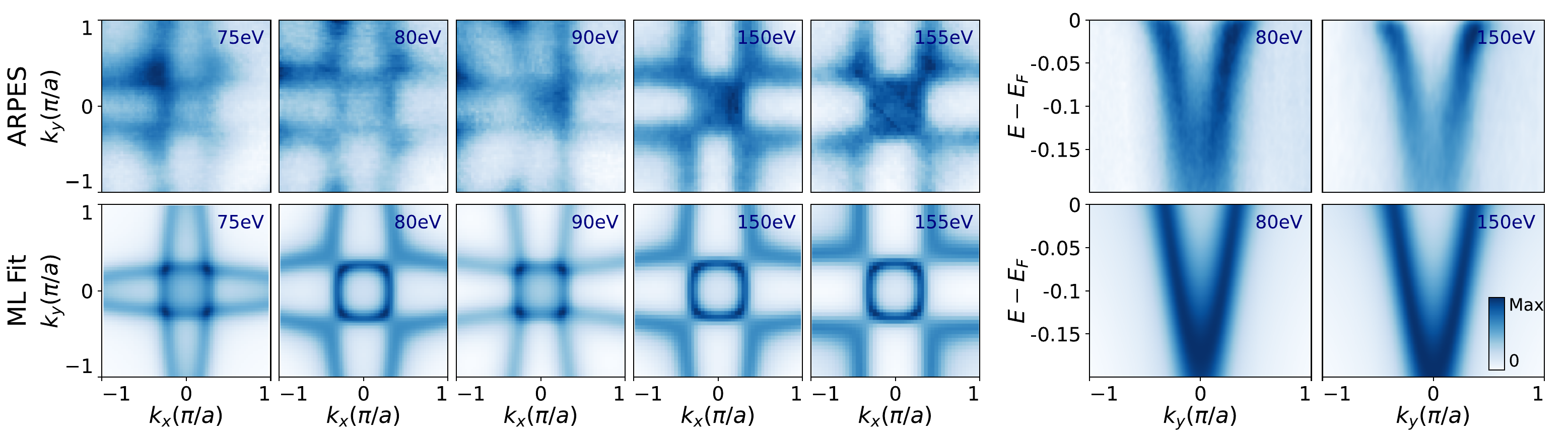}
    \caption{Fermi surface and $E_k$ dispersion for perovskite nickelates. Comparison between ARPES data (1$^{st}$ Row) and ML predictions (2$^{nd}$ Row) uses ML-fitted parameters with $t_1=0.31eV$, $t_2/t_1=0.06$, $\mu/t_1=1.95$. The blue values on each graph represent the photon energy used during measurement.}
    \label{NdNiO3}
\end{figure}


\subsection*{Comparison with experimental results}


To evaluate the performance of our method, we present a comprehensive comparison of the experimental data and the ML simulations in Figure 3. The first five columns show the Fermi surfaces, while the last two present the $E(\mathbf{k})$ dispersions. Using the ML-optimized parameters, we achieve good agreement with experimental observations for the overall Fermi surface topology, particularly around the $(\pi,\pi)$ region ($k_x = k_y = 1$) at photon energies of 150 eV and 155 eV. The corresponding weighted loss is $L = 0.042$. We want to emphasize that our simulation does not include matrix element effects, which can account for the intensity differences observed at symmetric $\mathbf{k}$-points in the experimental data. Since the waterfall feature is associated with many-body correlation effects that our TB model does not capture, we restrict the fitting of the \(E(\textbf{k})\) dispersion data to the energy range between \(E_F\) and \(-0.1\) eV, thereby excluding the waterfall region from the analysis. A comparison with a conventional fit of MDC peak positions on the same data is given in the Supplementary Materials. The two approaches yield comparable agreement with experiment.


\subsection*{\red{Dependence on sampling density}}
\begin{figure}[htbp]
    \centering
    \includegraphics[width=1.0\linewidth]{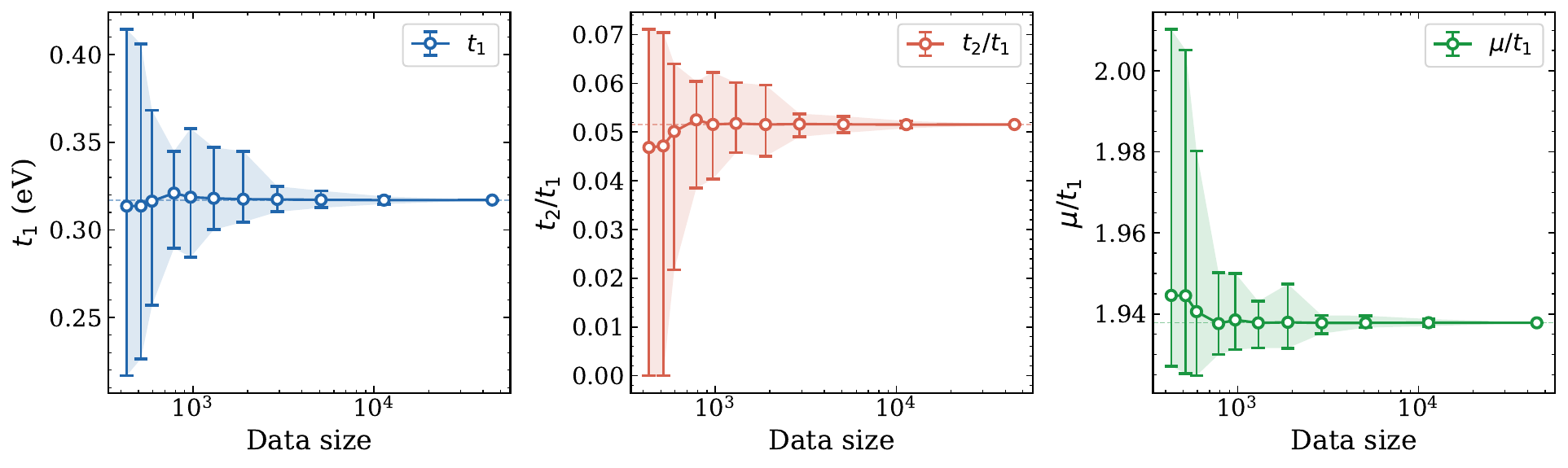}
    \caption{Convergence of the extracted tight-binding parameters with reduced momentum sampling density. The left, middle, and right panels show the fitted values of $t_1$, $t_2/t_1$, and $\mu/t_1$, respectively. Error bars indicate the standard deviation of the extracted parameters over the data subsets, while the markers represent the average values over all considered subsets.}
    \label{low_count}
\end{figure}
In ARPES experiments, a denser momentum-space sampling requires more acquisition time. To systematically learn the effect of the data density, we construct subsampled datasets from existing experimental measurements. Specifically, given an original $M \times N$ dataset and a sampling step $S$, we construct subsampled datasets using stride selection with offset $(O_x, O_y)$:
\[
I_{i,j} \rightarrow I_{O_x + iS,\; O_y + jS},
\]
where $O_x,O_y \in \{0,\dots,S-1\}$. The resulting grid size is
\[
M' = \left\lfloor \frac{M - O_x - 1}{S} \right\rfloor + 1, \quad
N' = \left\lfloor \frac{N - O_y - 1}{S} \right\rfloor + 1.
\]
This yields $S^2$ distinct subsamples.

We use all these subsamples to fit tight-binding parameters using our framework. As illustrated in Figure \ref{low_count}, at low resolution (large $S$), the fitted parameters exhibit significant fluctuations. As the resolution increases, the uncertainties decrease, and the fitted values gradually converge, demonstrating the stability of the fitting procedure. The convergent trends among the three parameters are consistent, and the values stabilize once approximately $3\times 10^3$ data points are included, which could provide guidance for future ARPES measurements.

\subsection*{Application to perovskite manganite}
\begin{figure}[htbp]
    \centering
    \includegraphics[width=1.0\linewidth]{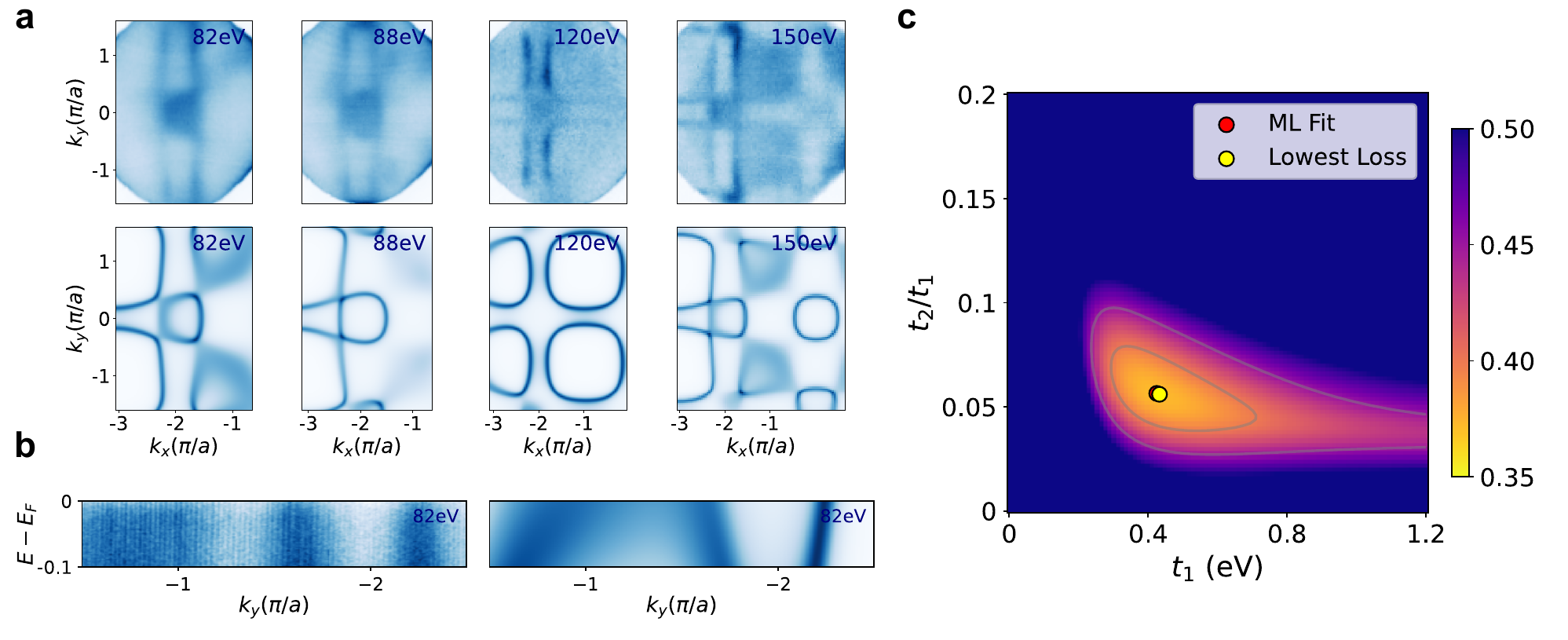}
    \caption{ \textbf{a}. La$_{1-x}$Sr$_x$MnO$_3$ ($x=0.33$) ARPES Fermi surface measurements (top row) and TB spectral intensity (bottom row) with parameters $t_1=0.42~eV$, $t_2/t_1=0.06$, $\mu/t_1=1.23$. 
    \textbf{b}. ARPES band dispersion (left panel) and ML-predicted band dispersion (right panel). The blue values in the upper right corner of each panel in \textbf{a} and \textbf{b} represent the photon energy used during ARPES measurements.
    \textbf{c}. Visualization of the loss landscape at $\mu = 1.23t_1$ with a 101$\times$101 grid for both $t_1$ and $t_2/t_1$. Both the Fermi surface and band dispersion ARPES data in \textbf{a} and \textbf{b} are utilized during the ML procedure. 
}
    \label{Manganites}
\end{figure}

More importantly, once the ML surrogate is well trained, the proposed framework can be easily applied to other materials that are described by the same tight-binding model. To demonstrate this generalization, we applied our approach to the manganite system La$_{1-x}$Sr$_x$MnO$_3$ with $x=0.33$ \cite{Millis1998}. As illustrated in Figure \ref{Manganites}{\bf b} and {\bf c}, our approach extracts tight-binding parameters without requiring experts' estimation. Notably, the entire process is completed within five minutes. The Fermi surfaces calculated with the extracted parameters reproduce the ARPES results across various photon energies quite well. The loss landscape in Figure \ref{Manganites}{\bf c} shows that the ML procedure achieves a minimal loss nearly coinciding with the global minimum, indicating that the predicted parameters ($t_1 = 0.42~eV$, $t_2/t_1 = 0.06$, and $\mu/t_1 = 1.23$) provide a globally optimal description of the combined Fermi-surface and band-dispersion data. The corresponding weighted loss is $L = 0.37$, which is much larger than that for the nickelate ($L = 0.042$), reflecting the broader spectral features of the manganite data.

\section*{Discussion}
Our work successfully bridges the gap between the complex experimental data and the theoretical models by using a SIREN-based neural network to simulate the electronic structure of perovskite nickelates and manganites. The agreement demonstrates that an intensity-based ML approach can automate the extraction of tight-binding parameters from high-dimensional ARPES data, from which the band structure can be evaluated on an arbitrarily fine grid. The primary advantages of this method are its efficiency, accuracy, and transferability. With an affordable initial training cost, the model can be directly applied to similar material systems described by the same tight-binding model without retraining. This technique offers a robust and standardized alternative to labor-intensive fitting procedures, opening a new avenue for the high-throughput analysis of electronic structures in materials science. We also expect our intensity-based approach to become more advantageous when more complex theoretical models are considered, including effects such as a momentum- and energy-dependent self-energy, electron-phonon coupling, and strong correlation effects, which are difficult to characterize by fitting peak positions alone.

Looking ahead, our framework points toward a broader integration of domain-specific deep learning with emerging advances in large language models. One direction is to combine models trained on spectroscopic data with more general reasoning tools, enabling AI systems to autonomously navigate model selection — for example, determining the relevant degrees of freedom, lattice symmetries, or matrix element terms needed for accurate electronic structure reconstruction. Such an integration could evolve into agent-like platforms that not only fit data, but also reason about the appropriate physical model and adapt to new material classes with minimal human input. Ultimately, this vision suggests a path toward autonomous discovery pipelines, where spectroscopic data across diverse quantum materials are interpreted in real time, providing unprecedented scalability and consistency in connecting experiments with theory.


\section*{Methods}
\label{Sec:Methods}

\subsection*{Sample preparation}
The training data were collected from a series of Nd$_{0.825}$Sr$_{0.175}$NiO$_3$ single-crystalline thin films synthesized via pulsed-laser deposition (PLD) on LSAT(001) substrates. The generalized example of La$_{0.67}$Sr$_{0.33}$MnO$_3$ thin films were grown by the same PLD technique on STO(001) substrates.

\subsection*{ARPES experiments}
The ARPES measurements were performed at the Stanford Synchrotron Radiation Lightsource (SSRL) beamline 5-2, equipped with a Scienta DA30 analyzer. To ensure optimal surface quality, thin films were annealed at \qty{500}{\degree C} for \qty{20}{min} under a base pressure of \num{1e-5} Torr of purified ozone. Linear-horizontally polarized light with photon energies ranging from \qtyrange{60}{180}{eV} was used to map the three-dimensional electronic structure. The base pressure during the ARPES measurements was better than \num{3e-11} Torr. All data were acquired at a base temperature of \qty{10}{K}. For Nd$_{1-x}$Sr$_x$NiO$_3$, the Fermi Surfaces were measured in the 1$^{st}$ Brillouin Zone (BZ) using incident photon energies of 70, 75, 80, 90, 150, 155eV, while band dispersions along X ($k_x=1, k_y=0$) to M ($k_x=1, k_y=1$) were recorded at 80 and 150eV. For La$_{1-x}$Sr$_x$MnO$_3$, the FS were measured at 82, 88, 120, and 150eV, with band dispersion measured at 82eV.

\subsection*{Tight-binding model}
The tight-binding Hamiltonian is given by:
\[
H_{tb} = -\sum_{ij} t_{ij}^{ab} c_{i a}^\dagger c_{j b} - \mu
\]
The hopping parameters are defined as:
\[
t_{i, i \pm \hat{\mu}}^{ab} = t_1 \, \phi_\mu^a \phi_\mu^b
\]
\[
t_{i, i \pm \hat{\mu} \pm \hat{\nu}}^{ab} = t_2 \left( \phi_\mu^a \phi_\nu^b + \phi_\mu^b \phi_\nu^a \right)
\]
where:
\[
\phi_x = \left(-\frac{1}{2}, \frac{\sqrt{3}}{2} \right), \quad
\phi_y = \left(-\frac{1}{2}, -\frac{\sqrt{3}}{2} \right), \quad
\phi_z = (1, 0)
\]
These correspond to the wave functions for $d_{3x^2 - r^2}$, $d_{3y^2 - r^2}$, and $d_{3z^2 - r^2}$ $\sigma$-bonding orbitals along the three axes. We diagonalize $H_{tb}$ to obtain the bare band dispersion for the system.

\subsection*{Data generation}
The experimental data include both the Fermi Surface and the E(k) dispersion. Based on the data range, we synthesize ARPES intensities in selected $\bf{k}$ ($k_x, k_y, k_z$) points, as shown in Table.\ref{dataset}. $k_x, k_y$ are renormalized and have units of $\pi/a$, where a is the lattice constant. $k_z$ is computed using the following equation:
\begin{equation}
    k_z=\sqrt{\frac{2m}{\hbar^2}(h\nu-\phi-E_B+V)-k_x^2-k_y^2}
\end{equation}
where $h \nu$ is the incident photon energy, $ \phi$ is the work function, $E_B$ is the binding energy and $V$ is the inner potential. We set $\phi=4.4eV, V=12eV$  and $\Sigma'=0, \Sigma''=100meV, \Delta k_z=0.2\frac{\pi}{a}$ for both materials. The dataset is generated using the parameters listed in Table 1, each of which is randomly and uniformly sampled from its specified range. The dataset is then divided into training, validation, and test sets in an 8:1:1 ratio.

\begin{table}[htbp]
\centering
\caption{Summary of the parameters used to generate the synthetic ARPES data.}
\label{dataset}
\begin{tabular}{ccccccc}
\toprule
Parameter & $k_x(\pi/a)$ & $k_y(\pi/a)$ & $h\nu(eV)$ & $t_1(eV)$  & $t_2/t_1$  & $\mu/t_1$ \\
\midrule
Range & [-3, 1] & [-1, 3] & [70, 160] & [0, 1.2] & [0, 0.2] & [0.5, 2.5]\\
\midrule
Quantity & 201 & 201 & 12 & 50 & 12 & 16\\
\bottomrule
\end{tabular}
\end{table}

\subsection*{SIREN model training}
The SIREN model was trained to predict spectral intensity by minimizing the Mean Squared Error (MSE) between $I_{ML}$ and $I_{TB}$. We use the Adam optimizer, batch size = 65536, initial learning rate = 0.001, which is decayed exponentially. The model was trained for 55 epochs on an NVIDIA B200 GPU. To accelerate training, only one-fifth of the training dataset was used in each epoch, resulting in an average epoch duration of approximately 150s.

\subsection*{Tight-binding parameters extraction}
The tight-binding parameters were optimized using the Adam algorithm with an initial learning rate = 0.01 and an exponential decay for 3000 iterations. The chemical potential $\mu$ is not an input of the network. An experimental data point at energy $E_{exp}$ relative to the Fermi level is evaluated at $E=E_{exp}+\mu$, so that the Fermi surface corresponds to $E=\mu$, and it's treated as a free parameter and optimized together with $t_1$ and $t_2$.

To integrate all experimental datasets, we employed a weighted loss function, where the weights were determined by the number of data points $N_i$ in each ARPES dataset. 
\begin{equation}
    L = \frac{1}{N_{all}}\sum_i L_iN_i
\end{equation}
In our simulation, the matrix element was not included, resulting in a symmetrized Fermi surface in the $1^ {st}$ Brilliouin Zone (BZ). For gradient-based optimization, we used 1/4 of the ARPES Fermi surface data in $1^{st}$ BZ and part of ($E-E_F>-0.1eV$) E(k) dispersion data. Table 2 and 3 summarize the experimental data used for parameter extraction, including the incident photon energy, cutting boundaries, and data size.

\begin{table}[htbp]
\centering

\caption{Nd$_{1-x}$Sr$_x$NiO$_3$ data for tight-binding parameters extraction.}
\label{dataset_paras}
\begin{minipage}{\textwidth}
\centering

\begin{minipage}[t]{0.7\textwidth}
\centering
\begin{tabular}{>{\centering\arraybackslash}m{1.8cm}
                >{\centering\arraybackslash}m{0.9cm}
                >{\centering\arraybackslash}m{0.9cm}
                >{\centering\arraybackslash}m{0.9cm}
                >{\centering\arraybackslash}m{0.9cm}
                >{\centering\arraybackslash}m{0.9cm}}
\toprule
\makecell{Photon \\ Energy (eV)} & 75 & 80 & 90 & 150 & 155 \\
\midrule
$k_x$  & [-1, 0] & [-1, 0] & [0, 1] & [-1, 0] & [-1, 0] \\
\midrule
$k_y$  & [0, 1] & [0, 1] & [0, 1] & [0, 1] & [0, 1] \\
\midrule
Size   & 1296 & 1225 & 1122 & 625 & 600 \\
\bottomrule
\end{tabular}
\end{minipage}

\vspace{0.5cm}
\begin{minipage}[t]{0.4\textwidth}
\centering
\begin{tabular}{>{\centering\arraybackslash}m{1.8cm}
                >{\centering\arraybackslash}m{1.2cm}
                >{\centering\arraybackslash}m{1.2cm}}
\toprule
\makecell{Photon \\ Energy (eV)} & 80 & 150 \\
\midrule
$k_y$ & [0, 1] & [0, 1] \\
\midrule
$E-E_F$ & [0, -0.1] & [0, -0.1] \\
\midrule
Size   & 29088 & 20300 \\
\bottomrule
\end{tabular}
\end{minipage}

\end{minipage}
\end{table}

\begin{table}[htbp]

\caption{La$_{1-x}$Sr$_x$MnO$_3$ data for tight-binding parameters extraction.}

\label{dataset_paras}

\centering
\begin{tabular}{@{}c c@{}}
\begin{tabular}{>{\centering\arraybackslash}m{1.8cm}
                >{\centering\arraybackslash}m{1.3cm}
                >{\centering\arraybackslash}m{1.3cm}
                >{\centering\arraybackslash}m{1.3cm}
                >{\centering\arraybackslash}m{1.3cm}}
\toprule
\makecell{Photon \\ Energy (eV)} &  82 & 88 & 120 & 150 \\
\midrule
$k_x$  & [-3, -1] & [-3, -1] & [-3, -0.5] & [-3, 0]  \\
\midrule
$k_y$  & [-1, 1] & [-1, 1] & [-1, 1] & [-1, 1]  \\
\midrule
Size   & 4623 & 4290 & 3864 & 3700  \\
\bottomrule
\end{tabular}

&

\begin{tabular}{>{\centering\arraybackslash}m{1.8cm}
                >{\centering\arraybackslash}m{1.5cm}}
\toprule
\makecell{Photon \\ Energy (eV)} & 82  \\
\midrule
$k_y$ & [-0.5, -2.5]  \\
\midrule
$E-E_F$ & [0, -0.1]  \\
\midrule
Size   & 56300 \\
\bottomrule
\end{tabular}
\end{tabular}

\end{table}

\section*{Data Availability}

The code used for analysis is available at \url{https://github.com/YuZ-TvT/ML-ARPES.git}. Data is available upon reasonable request.

\section*{Acknowledgments}

This work is supported by the U.S. Department of Energy, Office of Science, Basic Energy Sciences under Award No. DE-SC0022216, and Stanford Lab for Intelligent Matter (SLIM). The work at Stanford Institute of Materials and Energy Sciences (SIMES) and SLAC was also supported by the U.S. Department of Energy, Office of Basic Energy Sciences, Division of Materials Sciences and Engineering, under contract no. DE-AC02-76SF00515. The calculations are conducted at the University of Florida Research Computing.

\section*{Author Contribution}
The project was designed by C. J., Yong Z. and Z.-X. S.. Yong Z. performed the ARPES measurements. Yu Z. and Yong Z. performed the machine learning training with the help of C. J., H. T., S. L..
K. L., Y. L. and T. C. W. prepared the thin films under the advice of H.H.. The manuscript was written by Yu Z., Yong Z. and C.J. with input from all coauthors.

\section*{Competing Interest}

We have no conflict of interest to disclose.

\backmatter

\bibliography{main}

\end{document}